**Energy-Efficient Drone Logistics for Last-Mile Delivery: Implications of Payload-Dependent Routing Strategies**

**Ziyue Li, Ph.D. student**
Department of Civil and Environmental Engineering
Florida State University, Tallahassee, FL 32310
Email: lzy@eng.famu.fsu.edu

**Qianwen (Vivian) Guo, Ph.D.***
Assistant Professor
Department of Civil and Environmental Engineering
Florida State University, Tallahassee, FL 32310
Email: qguo@eng.famu.fsu.edu

**Paul Schonfeld, Ph.D.**
Professor
Department of Civil and Environmental Engineering
University of Maryland, College Park, MD 20742
Email: pschon@umd.edu

*Corresponding Author

## Abstract

Drone delivery is rapidly emerging as a cost-effective and energy-efficient alternative for last-mile delivery. Unlike ground vehicles, a drone's energy consumption depends on its payload in addition to travel distance. This creates a unique environmental challenge for multi-stop delivery tours, as the drone's total weight, and therefore its energy consumption rate, dynamically changes after each delivery. This paper investigates a novel green drone routing problem focused on maximizing energy efficiency. Through a series of motivating examples and numerical experiments, we demonstrate that energy-aware routing leads to several counter-intuitive routing strategies that contradict traditional distance-minimization delivery: a longer route may actually consume less energy than a shorter one; separate single-customer tours can be superior to a multi-stop tour; and a heterogeneous fleet, with drones of varying sizes, can achieve greater efficiency by matching drone capacity to specific delivery demands. In the numerical study, the green routing strategy shows energy savings in 67% of the instances. For these cases, the average energy saving is 2.17%, with a maximum saving of 5.97%, compared to minimum distance routing. These findings highlight the potential for green drone routing strategies to improve the sustainability of last-mile delivery.

## 1. Introduction

The world is warming faster than ever in recorded history, driven by greenhouse gas (GHG) emissions that trap the sun's heat. This phenomenon changes weather patterns, disrupts natural ecosystems, and poses significant risks to human beings (*United Nations, 2025*). In the United States, the transportation sector is a major contributor, accounting for 28% of the total GHG emissions (*U.S. Environmental Protection Agency, 2025*). In response, efforts over many decades have focused on shifting drivers from personal automobiles to public transit and freight from trucks to rail, as well as on introducing cleaner fuels (*Erdoğan & Miller-Hooks, 2012*).

As a complement to ground-based transportation, unmanned aerial vehicles (UAVs), or drones, have emerged as a popular option for package delivery in recent years. They offer numerous benefits, including lower operational costs, flexible route choices, reduced traffic congestion, and lower GHG emissions (*Doole et al., 2020*; *Stolaroff et al., 2018*; *Zhang et al., 2021*). Following Amazon's announcement of its Prime Air UAV in 2013, companies such as 7-Eleven, Alibaba, DHL, Domino, FedEx, and Google also began experimenting with drone delivery services (*Hossain, 2022*). Today, advancements in drone technology have significantly expanded their capabilities for last-mile delivery. For example, Walmart's delivery drones can now carry a payload of up to 10 pounds within a 6-mile radius (*Walmart, 2025*). Highlighting this trend, a 2021 report by ARK predicted that drones will be responsible for over 20% of package deliveries within the next five years (*Garg et al., 2023*).

The growing adoption of drones for last-mile delivery has led to a substantial increase in academic research, as reviewed by Garg et al. (*2023*). Many studies evaluate the efficiency, sustainability, and accessibility of drone-based delivery, with a specific focus on the environmental impacts. To quantify these impacts, energy consumption models have been developed to estimate GHG emissions for various drone delivery configurations (e.g., *Figliozzi, 2020*; *Kirschstein, 2020*). These studies highlight the substantial potential for GHG emission reductions through the integration of drones into last-mile logistics.

Compared to traditional vehicle routing, some unique characteristics of drones present new operational challenges (*Cheng et al., 2020*). A primary challenge is the drone's limited battery capacity, as its flight range is highly sensitive to payload, speed, and weather conditions (*Dorling et al., 2016*).[1] A drone's energy consumption, in particular, is highly dependent on its current payload (*Figliozzi, 2017*). This is critical for multi-stop delivery routes because as a drone serves each customer, its payload decreases, which dynamically changes its energy consumption rate throughout the trip. This is a key difference from truck delivery, where the packages' weight and fuel weight are often negligible compared to the vehicle's own weight, and payload-related energy fluctuations can be ignored. Although some studies consider these payload-dependent energy dynamics in their drone routing models (e.g., *Dorling et al., 2016*; *Cheng et al., 2020*; *Bruni et al., 2023*), the potential for routing strategies to be specifically optimized to minimize GHG emissions is still underexplored. Moreover, many studies only consider a homogeneous drone fleet where all drones are identical. This neglects the fact that different drones have different capabilities. Utilizing a variety of drone types can allow for better deployment strategies and lead to improved energy efficiency.

To address this research gap, our study proposes a novel green drone routing model for multi-visit deliveries using a heterogeneous drone fleet. The contributions of this work are twofold. First, we quantify how payload-dependent energy dynamics make green drone routing different from traditional truck routing. Second, we demonstrate that strategically deploying different drone types for different delivery scenarios can lower the total energy consumption and the resulting GHG emissions. We highlight three counter-intuitive phenomena:

---

[1] This paper assumes that drones are powered by electric batteries. Although other propulsion technologies (e.g., gas-powered and hybrid systems) are possible, our analysis is restricted to electric battery propulsion.

- **Distance vs. Energy**: A longer route can result in lower total energy consumption than the shortest route.
- **Tour Structure**: Separate single-customer tours can outperform consolidated multi-stop tours in terms of carbon efficiency.
- **Fleet Synergy**: A heterogeneous fleet by leveraging drones of varying sizes can achieve superior energy efficiency by matching drone capabilities to specific delivery demands.

It is important to emphasize that the primary objective of this study is not the development of new optimization algorithms. Instead, we focus on understanding the environmental implications of payload-dependent drone routing strategies. We aim to identify why drone routing logic diverges from classical distance-minimization routing, and to quantify the GHG emission reduction potential when shifting toward emission-aware objective functions. By focusing on these strategic insights rather than algorithmic speed, this work provides practical guidelines for drone routing strategies and fleet management policies in green logistics.

The paper is structured as follows. Section 2 reviews the relevant literature on drone energy consumption models and related routing problems. Section 3 formulates the green drone routing problem. Section 4 presents the results of numerical experiments and discusses their key implications. Finally, conclusions are drawn in Section 5.

## 2. Literature Review

While comprehensive reviews of drones in last-mile delivery and their use in vehicle routing problems can be found in Garg et al. (*2023*) and Rojas Viloria et al. (*2021*), respectively, this section focuses specifically on drone routing problems that explicitly consider energy consumption. Since these problems are built upon energy dynamics, we provide a detailed review of drone energy consumption models with technical formulas in **Appendix A** for interested readers.

A drone's energy consumption is not only a function of travel distance; it is also highly dependent on its payload. Furthermore, drones operate under battery and weight capacity limits, which restrict both the flight range and the amount of goods they can carry per trip. These factors distinguish drone routing from traditional vehicle routing, and consequently, specialized models or constraints are required to solve drone routing problems effectively.

A common approach in the literature is to impose strict assumptions to manage the problem's complexity. These typically include defining a limited service area and, most notably, restricting a drone to visit only one customer per trip (i.e., assuming unit capacity). By also assuming that a drone has sufficient battery to complete any single delivery, these models could bypass the complexities of energy consumption. This approach can be seen in studies of both drone-only deliveries (e.g., *Figliozzi, 2017, 2020*; *Goodchild & Toy, 2018*) and truck-drone deliveries (e.g., *Agatz et al., 2018*; *Carlsson & Song, 2018*; *Kang & Lee, 2021*; *Murray & Chu, 2015*; *Zhang et al., 2024*).

In contrast, another stream of research relaxes the unit capacity constraint and allows a drone to visit multiple customers per trip. However, to keep the models tractable, these studies often introduce other simplifications regarding energy consumption. For example, Choi and Schonfeld (*2017*) studied the optimization of multi-package deliveries by assuming homogeneous package weights. They set the battery capacity to be large enough for a drone to complete a round trip across the service area while carrying a single parcel. The objective of their model is to optimize the drone fleet size based on the relation among battery capacity, payload, and flight range. Poikonen and Golden (*2020*) investigated a multi-visit, truck-drone system that allows for packages of varying weights, but they make a critical assumption that a drone's battery life is a fixed duration, independent of its payload.

Recently, a growing body of research has started to explicitly incorporate energy consumption into drone routing models. For drone-only routing, the earliest work, to the best of our knowledge, is by Dorling et al. (2016). They addressed drone routing problems with payload-dependent energy consumption, using a linear function to approximate power consumed by a drone and a simulated annealing (SA) heuristic to find solutions. Cheng et al. (2020) formulated a multi-trip drone routing problem that included constraints for energy consumption and battery capacity. They proposed a branch-and-cut algorithm to solve the problem. A more recent study by Bruni et al. (2023) focused on tactical decisions about the selection of shared fulfillment centers used as the drone launch and retrieval stations and the fleet size. Their model explicitly considers the non-linear and load-dependent nature of the drone's energy consumption function.

For truck-drone routing problems, several studies have incorporated energy constraints as well. Masmoudi et al. (2022) addressed vehicle routing problems with drones equipped with multi-package compartments, allowing drones to visit multiple customers on a single trip. With truck-drone synchronization, a drone can return to a truck at a customer node to swap its battery and to pick up more packages from the current truck. They proposed an adaptive multi-start simulated annealing (AMS-SA) metaheuristic algorithm to solve the problem efficiently. Xia et al. (2023) considered a vehicle routing problem with load-dependent drones, explicitly incorporating energy constraints into their optimization model, which they solved using a branch-and-price-and-cut (BPC) algorithm. Jung (2025) used mixed-integer linear programming (MILP) for parcel delivery task allocation between UAVs and unmanned ground vehicles (UGVs). The model's optimization function for cost and time competitiveness accounted for factors such as the characteristics of the battery pack, allowable weight of delivery parcels, and corresponding electricity consumption. Jiang et al. (2025) studied a multi-visit vehicle routing problem with a heterogeneous drone fleet. They developed a MILP model that considers energy consumption and proposed a hybrid variable neighborhood search and simulated annealing (VNS-SA) algorithm to solve large-scale instances.

The study most similar to our work is by Lu et al. (2025), who proposed a green drone multi-package delivery routing problem (GDMPDRP) that minimizes total energy consumption. However, we identify three key limitations in their work. First, their model assumes that there are enough drones (i.e., not less than the number of customers), which is often unrealistic for practical operations. Second, it only considers a homogeneous drone fleet, overlooking the potential benefits of using different drone types. Third, the study presents a solution without discussing the strategic implications of green drone routing.

While many studies reviewed above, as summarized in **Table 1**, incorporate energy consumption as a constraint or as a component of the objective function, few treat energy minimization as the primary objective. Consequently, the strategic implications of minimizing energy consumption, and how this goal fundamentally changes routing decisions remain unclear. This gap motivates our study.

**Table 1 Drone-only or truck-drone routing problems with energy function.**

| Study | Objective function | Carrier type | Drone type | GHG reduction implications |
|---|---|---|---|---|
| Masmoudi et al. (2022) | Maximize total profit (revenue from delivery minus routing costs for trucks and drones) | Truck-drone | Homogeneous | ✗ |
| Xia et al. (2023) | Minimize total cost (fixed cost of trucks and travel cost of trucks and drones) | Truck-drone | Homogeneous | ✗ |

| Jung (*2025*) | Minimize total cost (battery pack cost, drone/truck rental cost, and energy cost) | Truck-drone | Homogeneous | ✗ |
|---|---|---|---|---|
| Jiang et al. (*2025*) | Minimize delivery completion times | Truck-drone | Heterogeneous | ✗ |
| Dorling et al. (*2016*) | Minimize the overall delivery time / total cost (drone usage cost and energy cost) | Drone-only | Homogeneous | ✗ |
| Cheng et al. (*2020*) | Minimize total cost (travel cost and energy cost) | Drone-only | Homogeneous | ✗ |
| Bruni et al. (*2023*) | Minimize total cost (delivery cost, drone usage cost, and tactical cost related to packing and handling services) | Drone-only | Homogeneous | ✗ |
| Lu et al. (*2025*) | Minimize energy consumption | Drone-only | Homogeneous | ✗ |
| Our study | Minimize energy consumption | Drone-only | Heterogeneous | ✓ |

It is important to distinguish our proposed model from existing "green" routing problems. The classic green vehicle routing problem (G-VRP), introduced by Erdoğan and Miller-Hooks (*2012*), is designed for fleets of Alternative Fuel Vehicles (AFVs), which run on greener fuel sources but have a shorter driving range and rely on limited refueling infrastructure. In that context, the objective is to minimize total travel distance. A more direct comparison is the green UAV routing problem (GUAVRP) proposed by Coelho et al. (*2017*), which optimizes multiple objectives simultaneously, such as minimizing total traveled distance and makespans of the last collected and delivered package. However, while GUAVRP considers battery constraints, it does not explicitly model the dynamics of energy consumption during the routing process. Our problem is different from both. Unlike G-VRP, where energy consumption is directly proportional to distance, a drone's energy use is influenced by both distance and varying payload weight, making the relation far more complex. Unlike GUAVRP, our model explicitly captures the dynamics of payload-dependent energy consumption and treats minimizing this factor as the core objective, rather than as one of several competing goals.

Finally, while a significant portion of the existing literature focuses on computational and algorithmic speed to solve large scale instances, the aim of this work is fundamentally different. Our research seeks to identify why and how drone routing differs from classical distance-minimization models to achieve the true GHG emission minimization objective.

## 3. Green Drone Routing Problem

### 3.1 Notation

Let the depot be represented by node 0 and the customers by nodes $1, 2, \cdots n$, where $n$ is the total number of customers. The set of customers is defined as $N = \{1, 2, \cdots, n\}$. The distance between any two nodes $i$ and $j$ is denoted by $d_{ij}$. For each customer $i \in N$, the corresponding package has a weight of $m_i$.[2] The delivery information for all customers is assumed to be known in advance.

---

[2] Strictly speaking, all quantities denoted by $m$ and $w$ in this paper represent mass (kg), rather than force (N), where mass = weight/gravity constant. Throughout the paper, we use the terms *weight* and *mass* interchangeably to remain consistent with the terminology commonly used in the literature (e.g., *tare mass*, *self-weight*). For clarity, $m$ denotes **parameters** such as the drone's self-weight or package weight, which are fixed, while $w$ denotes **variables** such as the drone's total weight during a routing trip, which changes dynamically. All units refer to mass in kilograms.

In this paper, a *flight* refers to a drone's movement from a single takeoff to a single landing. A *tour* is a complete route where a drone departs the depot, visits one or more customers, and returns to the same depot.

For simplicity, this study assumes that the energy consumption for each flight segment is proportional to the drone's total weight $w$ during that segment, which stays constant because the battery and payload weights do not change within the segment. This assumption is supported by the existing literature in **Appendix A**. Based on this, the energy consumption $E_{ij}$ for a single flight from node $i$ to node $j$ is calculated with the following linear model:

$$E_{ij} = (e_l + e_f d_{ij})w, \tag{1}$$

where $e_l$ is the energy consumption coefficient for takeoff, hovering, and landing per unit of weight, while $e_f$ is the energy consumption coefficient for level flight per unit of distance and per unit of weight. A drone's total weight, $w$, is the sum of two components: its self-weight, $m_0$, which includes the drone's tare weight and battery, and its current payload. The payload is defined as the total weight of all packages for customers who have not yet been visited on a given tour.

In the proposed green drone routing problem in Section 3.3, we consider a heterogeneous fleet consisting of $K$ different types of drones, indexed by the set $\mathcal{K} = \{1,2,\cdots,K\}$. Each drone type $k \in \mathcal{K}$ is characterized by distinct energy consumption coefficients: $e_{lk}$ for takeoff, hovering, and landing, and $e_{fk}$ for level flight. Its self-weight is $m_{0k}$, while its maximum total weight and energy capacity are $W_k$ and $E_k$, respectively. The fleet is composed of a fixed number of drones of each type. Let $h_k$ be the number of available drones of type $k$, where $H_k = \{1,2,\cdots,h_k\}$ is the index set for these specific drones.

A complete notation list is provided in **Table 2**.

**Table 2 Notational glossary.**

| | |
|---|---|
| *Notations used in the green drone routing problem (G-DRP) in Section 3.3* | |
| **Sets** | |
| $N$ | The set of customers |
| $\mathcal{K}$ | The set of types of drones |
| $H_k$ | The index set of drones of type $k$ |
| **Indices** | |
| $0$ | Depot |
| $i,j$ | Customer |
| $k$ | Drone type |
| $t$ | Drone index |
| **Parameters** | |
| $n$ | The total number of customers |
| $K$ | The number of different types of drones |
| $h_k$ | The number of available drones of type $k$ |
| $e_{lk}$ | Energy consumption coefficient for takeoff, hovering, and landing of type $k$ drone |
| $e_{fk}$ | Energy consumption coefficient for level flight of type $k$ drone |
| $d_{ij}$ | The distance between $i$ and $j$ |
| $m_i$ | The package weight of customer $i$ |
| $m_{0k}$ | The drone's self-weight of type $k$ (including the drone's tare mass and battery mass) |
| $W_k$ | The drone's maximum total weight of type $k$ |
| $E_k$ | The drone's maximum energy capacity of type $k$ |
| **Variables** | |
| $x_{ijkt}$ | Whether a drone of type $k$ on tour $t$ travels from node $i$ to $j$ |

| $w_{ikt}$ | The drone's total weight upon departure from node $i$ in the tour $t$ by type $k$ |
|---|---|
| $u_{ikt}$ | The sequence position of customer $i$ in the tour $t$ by type $k$ drone |
| $E$ | Total energy consumption |
| *Other notations used in Section 3.2* | |
| $m_0, m_0^L, m_0^S$ | Self-weight (including battery) of a drone, a large drone, and a small drone |
| $m$ | Payload mass |
| $e_l$ | Energy consumption coefficient for takeoff, hovering, and landing |
| $e_l^L, e_l^S$ | Energy consumption coefficient for takeoff, hovering, and landing of a large drone and a small drone |
| $e_f$ | Energy consumption coefficient for level flight |
| $e_f^L, e_f^S$ | Energy consumption coefficient for level flight of a large drone and a small drone |
| $E_1, E_2, E^L, E^S$ | Total energy consumption of route 1, route 2, large-drone, and small-drone |
| $\Delta E$ | Energy consumption difference between two routes |
| $D_1, D_2$ | Total travel distances of route 1 and route 2 |
| $d$ | Distance of a flight |

## 3.2 Motivating examples

To help readers develop an intuition for the key differences between green drone routing and traditional vehicle routing, this subsection presents three motivating examples. For clarity, we omit units and use the following nominal values for the computations: the energy coefficients are set to $e_l = 1$ and $e_f = 1$, and the drone's self-weight is $m_0 = 10$. These simplified values are intended only to provide an intuitive understanding of the core concepts. The formal numerical experiments in Section 4 use realistic parameters.

### 3.2.1   Example 1

This example demonstrates a core principle of green drone routing: the shortest path is not always the most energy-efficient path. Because a package's weight can be a significant fraction of the drone's total weight, delivering heavier items first can reduce overall energy consumption, even if it requires a longer route. Consider the scenario in **Figure 1(a)**, where the drone's self-weight is $m_0 = 10$ and it must deliver three packages with weights $m_1 = 1$, $m_2 = 10$, and $m_3 = 1$.

First, consider the shortest-distance tour, $0 \rightarrow 1 \rightarrow 2 \rightarrow 3 \rightarrow 0$. The total distance is $D_1 = 4$. The total energy consumption is the sum of the energy for each flight, which is calculated to be $E_1 = (1 + 1) \times 22 + (1 + 1) \times 21 + (1 + 1) \times 11 + (1 + 1) \times 10 = 128$. Now, consider a tour that delivers the heaviest package first, $0 \rightarrow 2 \rightarrow 1 \rightarrow 3 \rightarrow 0$. This route is longer, with a total distance of $D_2 = 2 + 2\sqrt{2} = 4.83$. However, its total energy consumption is lower, calculated as $E_2 = (1 + \sqrt{2}) \times 22 + (1 + 1) \times 12 + (1 + \sqrt{2}) \times 11 + (1 + 1) \times 10 = 77 + 33\sqrt{2} = 123.67$.

Although the first route is shorter ($D_1 < D_2$), the second route consumes less energy ($E_2 < E_1$). This illustrates that prioritizing heavier goods can lead to energy savings. Furthermore, this choice can affect feasibility. If the drone had an energy limit of 125, the shortest-distance tour would be infeasible, while the energy-optimal "green" tour would be possible.

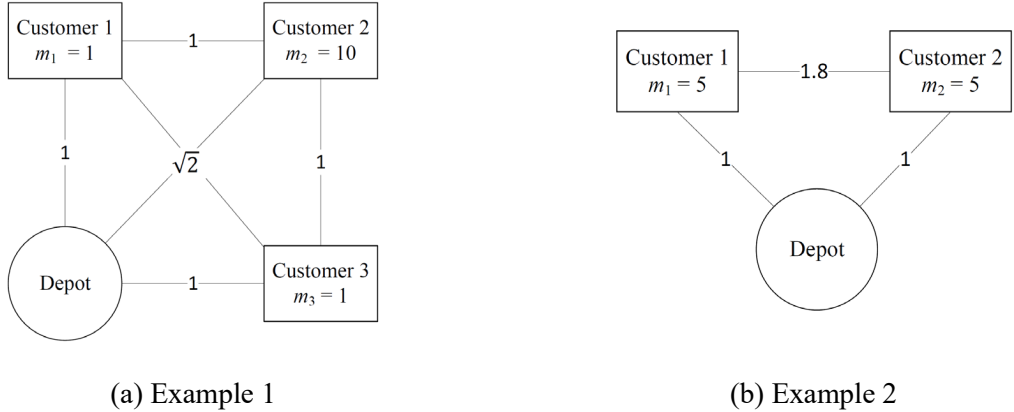

(a) Example 1          (b) Example 2

**Figure 1 Motivating examples.**

### 3.2.2   Example 2

This example demonstrates that a single multi-stop tour is not always more energy-efficient than making multiple separate deliveries. When customers are located far from each other, the energy required for the flight between them can be large. Consider the scenario in **Figure 1(b)**, where the drone's self-weight is $m_0 = 10$ and it must deliver two packages, each with a weight of 5 (i.e., $m_1 = 5, m_2 = 5$).

First, consider performing a single tour to visit both customers, following the route $0 \rightarrow 1 \rightarrow 2 \rightarrow 0$, This tour has a total distance of $D_1 = 3.8$, and its total energy consumption is calculated to be $E_1 = (1 + 1) \times 20 + (1 + 1.8) \times 15 + (1 + 1) \times 10 = 102$. Alternatively, the drone could make two separate deliveries. The total distance for these two trips $(0 \rightarrow 1 \rightarrow 0$ and $0 \rightarrow 2 \rightarrow 0)$ is $D_2 = 4$. However, the combined energy consumption for both trips is lower, calculated as $E_2 = 2 \times [(1 + 1) \times 15 + (1 + 1) \times 10] = 100$.

This counter-intuitive result, where the shorter tour consumes more energy ($D_1 < D_2$ but $E_1 > E_2$), occurs because the single tour requires a long flight between customers 1 and 2 while the drone is still carrying a payload. In such cases, making separate deliveries, despite increasing the total travel distance, is the more energy-efficient "green" strategy.

### 3.2.3   Example 3

This final example illustrates the benefits of using a heterogeneous fleet composed of large and small drones. We consider a large drone (denoted by superscript $L$) and a small drone ($S$) for a single round-trip delivery of a package with mass $m$ over a distance $d$. The self-weights of the large and small drones are $m_0^L$ and $m_0^S$, respectively. The key trade-off is that the large drone has a higher energy coefficient for takeoff, hovering, and landing ($e_l^L > e_l^S$) but a lower energy coefficient for level flight ($e_f^L < e_f^S$).[3] The round-trip energy consumption for each drone is given by:

$$E^L = \left(e_l^L + e_f^L d\right)\left(2m_0^L + m\right), \tag{2}$$

---

[3] This is supported by data reported in Tables 6 and 7 of Zhang et al. (*2021*). They provide the energy consumption per meter of travel (*Epm*) for a small drone with a total weight of 2.57 kg (tare mass 1.07 kg, battery mass 1kg, payload mass 0.5 kg) and for a large drone with a total weight of 24 kg (tare mass 7 kg, battery mass 10 kg, payload mass 7 kg). We focus on two linear models for steady level flight, LD and R2 (see *Zhang et al., 2021*, Table 5). Dividing *Epm* by the corresponding total weight yields the parameters $e_f^L$ and $e_f^S$. Calculations show that $e_f^S$ is 4.20 (LD) and 10.97 (R2), while $e_f^L$ is 3.99 (LD) and 8.81 (R2). These results support the assumption that $e_f^L < e_f^S$.

$$E^S = \left(e_i^S + e_f^S d\right)\left(2m_0^S + m\right). \tag{3}$$

Eq. (2) accounts for two flights: the large drone departs the depot with a total weight of $m_0^L + m$ and returns with a total weight of $m_0^L$. A similar calculation applies to Eq. (3).

To determine which drone is more efficient, we analyze the energy difference, $\Delta E = E^L - E^S$:

$$\begin{aligned}
\Delta E = E^L - E^S &= 2\left(e_i^L m_0^L - e_i^S m_0^S\right) + 2\left(e_f^L m_0^L - e_f^S m_0^S\right)d \\
&\quad + \left(e_i^L - e_i^S + \left(e_f^L - e_f^S\right)d\right)m.
\end{aligned} \tag{4}$$

The sign of $\Delta E$ depends on both the delivery distance $d$ and the payload mass $m$. First, for deliveries with a sufficiently large payload $m$, the last term associated with the payload will become negative when the distance $d$ exceeds a certain threshold, making the large drone more efficient ($\Delta E < 0$). Second, for very short distances ($d \to 0$), the energy difference is dominated by the higher fixed costs of the larger drone's takeoff, hovering, and landing ($\Delta E \to e_i^L\left(2m_0^L + m\right) - e_i^S\left(2m_0^S + m\right), e_i^L > e_i^S, 2m_0^L + m > 2m_0^S + m$), making the small drone more efficient ($\Delta E > 0$).

Therefore, this trade-off leads to a conclusion that larger drones are more energy-efficient for long-distance and heavy-payload deliveries, while smaller drones are better for short-distance and light-payload deliveries. This demonstrates that a heterogeneous fleet, where the most suitable drone can be chosen for particular deliveries, can achieve overall energy savings compared to a homogeneous fleet (*Kang & Lee, 2021*).

## 3.3 Problem formulation

The Green Drone Routing Problem (G-DRP) can be formulated as follows:

$$(\text{G}-\text{DRP}) \quad \min \quad E = \sum_{i=0}^{n}\sum_{j=0}^{n}\sum_{k=1}^{K}\sum_{t=1}^{h_k}\left(e_{lk} + e_{fk}d_{ij}\right)x_{ijkt}w_{ikt}, \tag{5}$$

$$\text{s.t.} \quad \sum_{i=0}^{n}\sum_{k=1}^{K}\sum_{t=1}^{h_k}x_{ijkt} = 1, \qquad \forall j \in N, \tag{6}$$

$$\sum_{j=0}^{n}\sum_{k=1}^{K}\sum_{t=1}^{h_k}x_{ijkt} = 1, \qquad \forall i \in N, \tag{7}$$

$$\sum_{j=1}^{n}x_{0jkt} \le 1, \qquad \forall k \in \mathcal{K}, t \in H_k, \tag{8}$$

$$\sum_{i=1}^{n}x_{i0kt} \le 1, \qquad \forall k \in \mathcal{K}, t \in H_k, \tag{9}$$

$$x_{ijkt} \le \sum_{j'=1}^{n}x_{0j'kt}, \qquad \forall i \in N, j \in N \cup \{0\}, k \in \mathcal{K}, t \in H_k, \tag{10}$$

$$w_{0kt} = m_{0k} + \sum_{i=1}^{n}\sum_{j=0}^{n}m_i x_{ijkt}, \qquad \forall k \in \mathcal{K}, t \in H_k, \tag{11}$$

$$x_{ijkt}\left(w_{ikt} - w_{jkt} - m_j\right) = 0, \qquad \forall i \in N \cup \{0\}, j \in N, k \in \mathcal{K}, t \in H_k, \tag{12}$$

$$u_{ikt} - u_{jkt} + x_{ijkt}\sum_{i'=0}^{n}\sum_{j'=1}^{n}x_{i'j'kt} \le \sum_{i'=0}^{n}\sum_{j'=1}^{n}x_{i'j'kt} - 1, \qquad \forall i,j \in N, \forall k \in \mathcal{K}, t \in H_k, \tag{13}$$

$$w_{0kt} \le W_k, \qquad \forall k \in \mathcal{K}, t \in H_k, \tag{14}$$

$$\sum_{i=0}^{n}\sum_{j=0}^{n}(e_{lk}+e_{fk}d_{ij})x_{ijkt}w_{ikt} \le E_k, \qquad \forall k \in \mathcal{K}, t \in H_k, \tag{15}$$

$$x_{iikt}=0, \qquad \forall i \in N \cup \{0\}, \forall k \in \mathcal{K}, t \in H_k, \tag{16}$$

$$x_{ijkt} \in \{0,1\}, \qquad \forall i, j \in N \cup \{0\}, \forall k \in \mathcal{K}, t \in H_k, \tag{17}$$

where $x_{ijkt}$ is the decision variable, which is 1 if a drone of type $k$ on tour $t$ travels from node $i$ to $j$; $w_{ikt}$ is the drone's total weight upon departure from node $i$; and $u_{ikt}$ is the sequence position of customer $i$ in the tour.

The objective function (5) minimizes the total energy consumption across all drone tours. Constraints (6) and (7) ensure that each customer is visited exactly once by a drone. Constraints (8) and (9) state that each drone tour can be used at most once. The inequality ($\le$) is used because not all drones in the fleet must be dispatched; it is permissible for a drone to remain at the depot. Constraint (10) ensures that any flight leg flown by drone $k$ on tour $t$ can only be active ($x_{ijkt}=1$) if that specific tour has been initiated by the drone leaving the depot. Constraint (11) defines the initial weight of a drone tour, while (12) dynamically updates the drone's weight after each delivery. Constraint (13) is the Miller-Tucker-Zemlin (MTZ) formulation to prevent subtours (*Miller et al., 1960*). This ensures that if a drone travels from node $i$ to node $j$ (i.e., $x_{ijkt}=1$), then the position of $i$ ($u_{ikt}$) must precede the position of $j$ ($u_{jkt}$). We use the number of customers visited on a specific tour ($\sum_{i'=0}^{n}\sum_{j'=1}^{n}x_{i'j'kt}$) to provide a tighter bound. The maximum total weight and battery energy capacity for each drone type are enforced by constraints (14) and (15), respectively. Constraint (16) prevents self-loops. Finally, constraint (17) defines the domain of the decision variable.

The G-DRP defined by (5) – (17) can be reformulated into a MILP. This reformulation is provided in **Appendix B**. To ensure the model's applicability to real-world operations, several extensions, including no-fly zones, volume limit for payload, weather conditions, battery replacement, and delivery with time window, are detailed in **Appendix C**.

## 4. Numerical Experiments

The optimization model was implemented in GAMS (Distribution 45.7.0) and solved using the CPLEX 22.1 solver. All numerical experiments are conducted on a desktop computer with an Intel i7-13700 2.1 GHz CPU and 32 GB of RAM.

In Section 4.1, we conduct numerical experiments on randomly generated instances to analyze the fundamental behavior of the model. In Section 4.2, we test the widely used Solomon dataset (*Solomon, 1987*), which further demonstrates the potential for GHG emission reductions when using the green routing strategy compared to a traditional distance-minimization strategy.

### 4.1 Randomly generated instances

In this subsection, customer locations for the baseline scenario are randomly generated within a 10 km × 10 km square service area, with the depot located at the center (0,0). The package weight for each customer is drawn from a uniform distribution between 0.5 kg and 2.0 kg. Travel distances between all node pairs ($d_{ij}$) are calculated as Euclidean distances. Details of the generated instances are provided in **Appendix D**.

#### 4.1.1 Numerical inputs

The details of the baseline parameters used for numerical experiments are summarized in **Table 3**. We consider a heterogeneous fleet with two drone types: a small drone ($k=1$) and a large drone ($k=2$). The parameters for the small drone are adapted from Cheng et al. (*2018, 2020*) and Dorling et al. (*2016*), assuming a level flight speed of 21.6 km/h (6 m/s). The parameters for the large drone are based on Zhang et al. (*2021*) and are set to maintain the trade-off established in **Example 3**, where the large drone has a higher takeoff, hovering, and landing energy consumption coefficient but a lower level flight energy consumption coefficient ($e_{l1} < e_{l2}$ and $e_{f1} > e_{f2}$).

**Table 3 Baseline parameter values.**

| Notations | Definitions | Baseline values | Units |
|---|---|---|---|
| $n$ | Number of customers | 10 | --- |
| $K$ | Number of drone types | 2 | --- |
| $h_k$ | Number of available drones of type $k$ | $h_1 = 3, h_2 = 2$ | |
| $m_{0k}$ | Self-weight of drone type $k$ | $m_1 = 3, m_2 = 10$ | kg |
| $e_{lk}$ | Takeoff, hovering, and landing energy coefficient for type $k$ | $e_{l1} = 0.3, e_{l2} = 1$ | Wh/kg |
| $e_{fk}$ | Level flight energy coefficient for type $k$ | $e_{f1} = 10, e_{f2} = 5$ | Wh/(kg×km) |
| $E_k$ | Battery energy capacity of drone type $k$ | $E_1 = 300, E_2 = 1,500$ | Wh |
| $W_k$ | Maximum total weight of drone type $k$ | $W_1 = 5, W_2 = 20$ | kg |

### 4.1.2 Routing solutions

The optimal solution for the baseline scenario is presented in **Figure 2**. The results show a clear task division between the two drone types, consistent with the principles demonstrated in the motivating examples in Section 3.2.

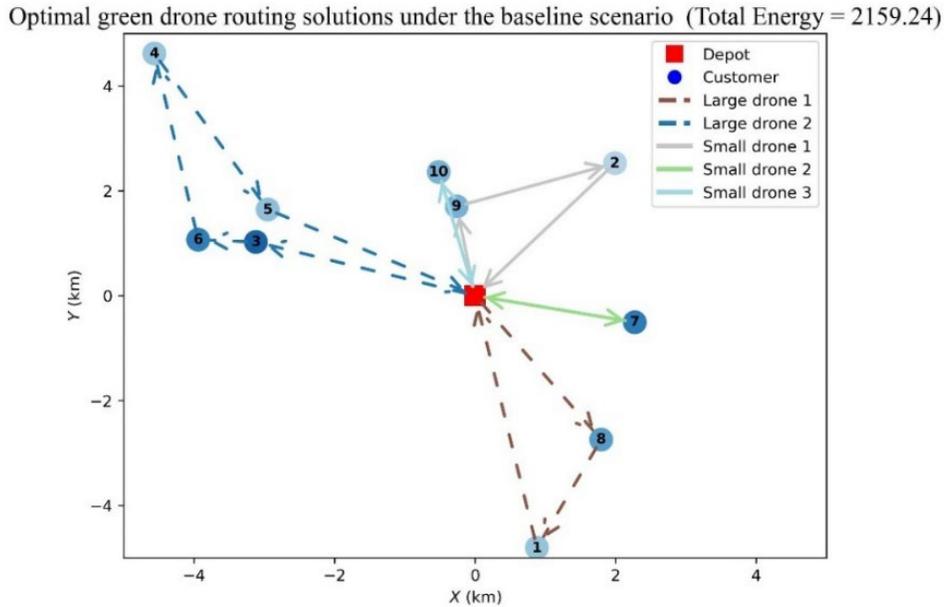

**Figure 2 Optimal green drone routing solution for the baseline scenario. Darker nodes indicate heavier package weights.**

Small drones (solid lines) are primarily dispatched for tours with fewer stops and lighter overall payloads. For instance, small drone 1 serves two nearby customers (2 and 9), while others handle single-customer deliveries (i.e., small drone 2 serves customer 7, and small drone 3 serves customer 10). Large drones (dashed lines), with their greater capacity, are used for longer, multi-stop tours with heavier cumulative loads and to reach distant customers. For example, large drone 1 serves customers 1 and 8, while large drone 2 serves customers 3, 4, 5, and 6. Notably, customers located at the edge of the service area, such as customer 1 and customer 4, are exclusively served by large drones in this solution.

This confirms the conclusion from **Example 3**: small drones are more efficient for less demanding tours, while large drones are better suited for long-distance and multi-package deliveries. This is consistent with the findings in Kang and Lee (*2021*) and Luo et al. (*2024*).

An interesting observation is that, although customers 9 and 10 are geographically close, they are served by two different small drones on separate tours. This is because a single small drone visiting both customers in one tour would be infeasible, as the total weights of 5.14 kg exceeds the maximum allowable weight of 5 kg.

### 4.1.3    Sensitivity analyses

First, we evaluate the benefits of a heterogeneous fleet by comparing the baseline solution to a scenario where only large drones are available. A small-drones-only scenario is found to be infeasible for this set of deliveries.

The optimal solution using only large drones is shown in **Figure 3**. While the routing for most customers remains similar to the baseline solution in **Figure 2**, the assignments for customers 2, 9, and 10 change. In the heterogeneous solution, these customers are efficiently served by two small drones, with one visiting customers 2 and 9 while the other visits customer 10. In the large-drones-only solution, two large drones are dispatched to serve them, with one visiting customer 2 and the other visiting customers 9 and 10.

Importantly, the total energy consumption for the large-drones-only scenario (2460.74 Wh) is higher than for the heterogeneous baseline scenario (2159.24 Wh). This demonstrates that while large drones have greater capacity, forcing them to handle smaller, short-distance deliveries is inefficient. A heterogeneous fleet reduces overall energy consumption by assigning the most appropriately sized drone to each delivery task.

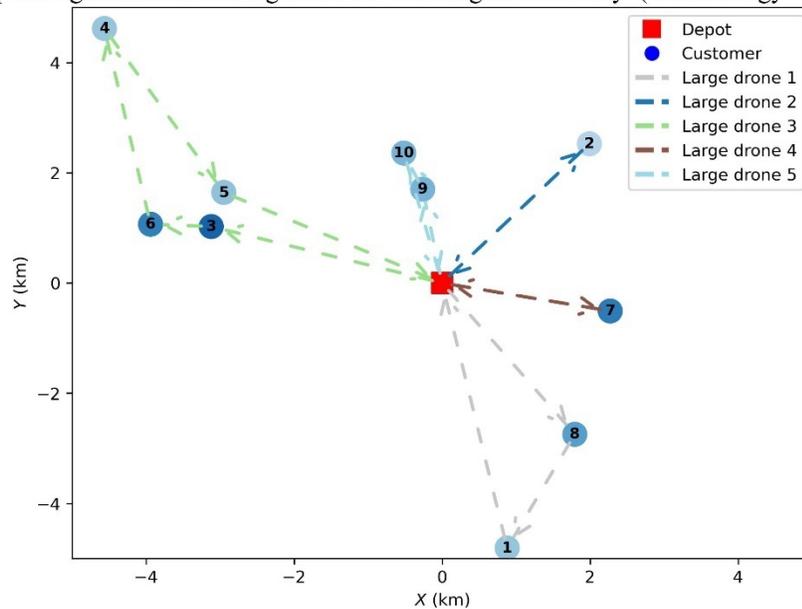

**Figure 3 Optimal solutions for the large-drones-only scenario.**

Second, we vary the energy parameters for the large drone ($e_{l2}$ and $e_{f2}$), as these values are adapted from the literature and not empirically tested for our specific case. **Figure 4** presents the total energy consumption and the resulting routes for various parameter values. As expected, the total energy consumption increases as the energy coefficients increase. More interestingly, the optimal routing solution remains the same across most scenarios, but changes in three specific cases where both $e_{l2}$ and $e_{f2}$ are relatively low ($e_{l2} = 0.5, e_{f2} = 4$; $e_{l2} = 0.5, e_{f2} = 5$; $e_{l2} = 1, e_{f2} = 4$). In these specific cases, the large drone becomes so energy-efficient that it is optimal for it to take a detour to serve customer 7, a customer previously served by a small drone in the baseline scenario. This highlights that the routing decisions are influenced by the energy efficiency of the drones in the fleet.

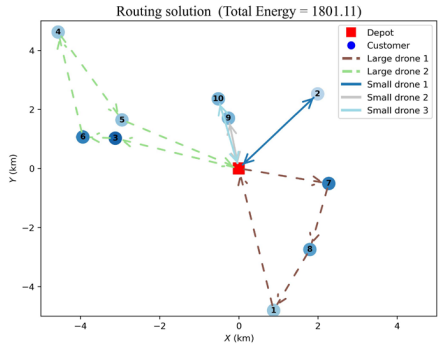
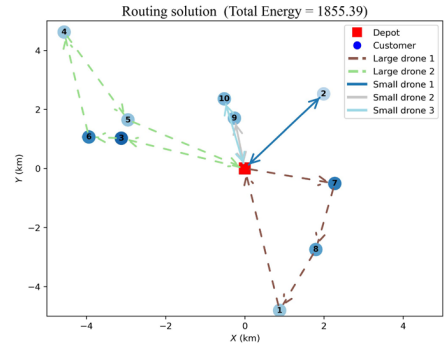
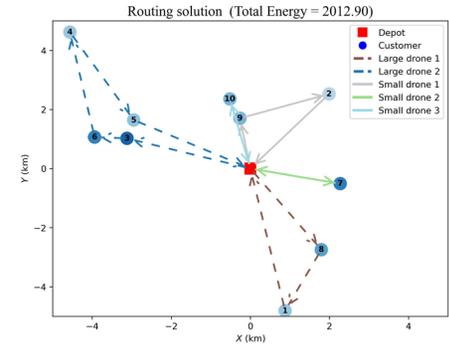

$e_{f2}$   4

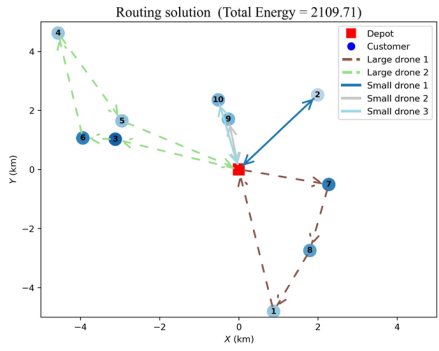
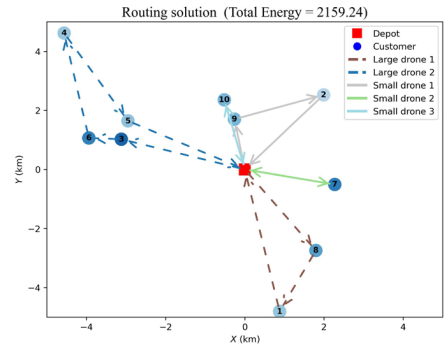
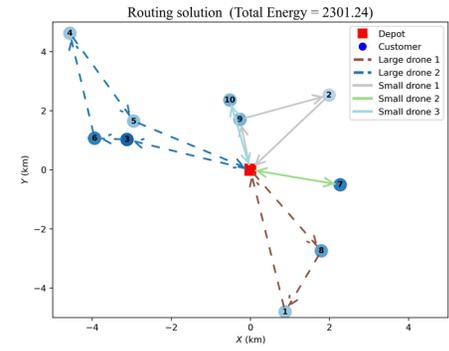

$e_{f2}$   5

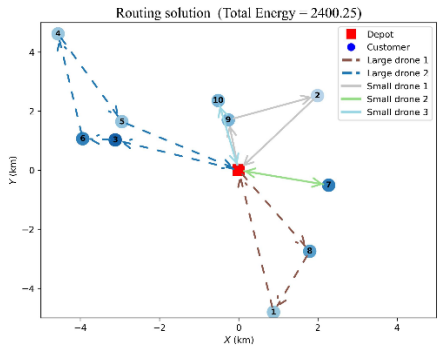
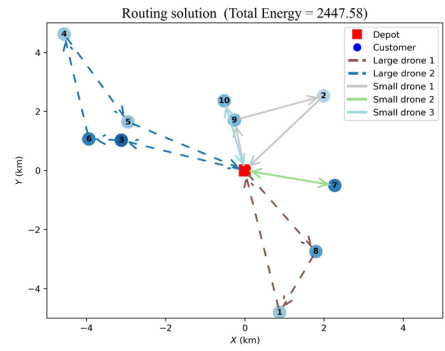
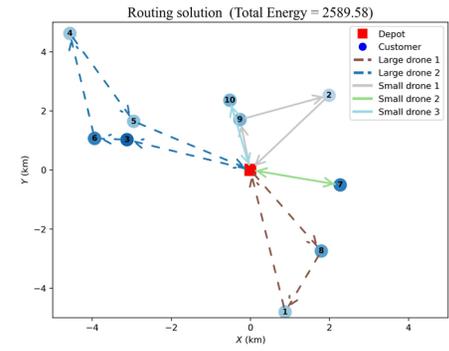

0.5                    1                    2.5

$e_{l2}$

1          **Figure 4 Sensitivity analysis of large drone energy parameters. The baseline scenario is located at the center of the plot.**

Third, we conduct a sensitivity analysis on package weights by rescaling the baseline weights to fit three different scenarios: (1) Light weights, where weights are uniformly distributed between 0.5 kg and 1.0 kg; (2) Heavy weights, with weights between 1.5 kg and 2.0 kg; and (3) Dispersed weights, with weights between 0.25 kg and 3.0 kg. The results are shown in **Figure 5**.

The analysis reveals that the distribution of package weights directly impacts the optimal routing strategy, even when the same drone serves the same set of customers. For example, a large drone serves customers 3, 4, 5, 6 in both baseline (a) and heavy-weight (c) scenarios. However, when the package weights increase, the optimal sequence of visits changes. For large drone 2, calculations show that the total distances of the tours $0 \rightarrow 3 \rightarrow 6 \rightarrow 4 \rightarrow 5 \rightarrow 0$ and $0 \rightarrow 5 \rightarrow 3 \rightarrow 6 \rightarrow 4 \rightarrow 0$ are 14.47 km and 14.94 km, respectively. In scenario (c), the drone selects the longer route, yet it consumes less energy, consistent with the findings illustrated in **Example 1**.

Furthermore, heavier package weights discourage multi-stop tours for smaller drones. In the baseline (a) and light-weight (b) scenarios, at least one small drone serves two customers. However, in the heavy-weight (c) and dispersed-weight (d) scenarios, all small drones are restricted to single-customer tours.

Meanwhile, under scenario (b), customers 2 and 7 are served by two small drones. Although the combined weight of serving both customers with a single small drone is 4.45 kg, below the drone's weight capacity, the energy consumption would be higher than with separate deliveries. This validates the result from **Example 2**, showing that separate deliveries may consume less energy than a single combined delivery.

Additionally, as expected, total energy consumption is sensitive to package weight, with the heavy-weight (c) scenario requiring 15.5% more energy than the light-weight (b) scenario.

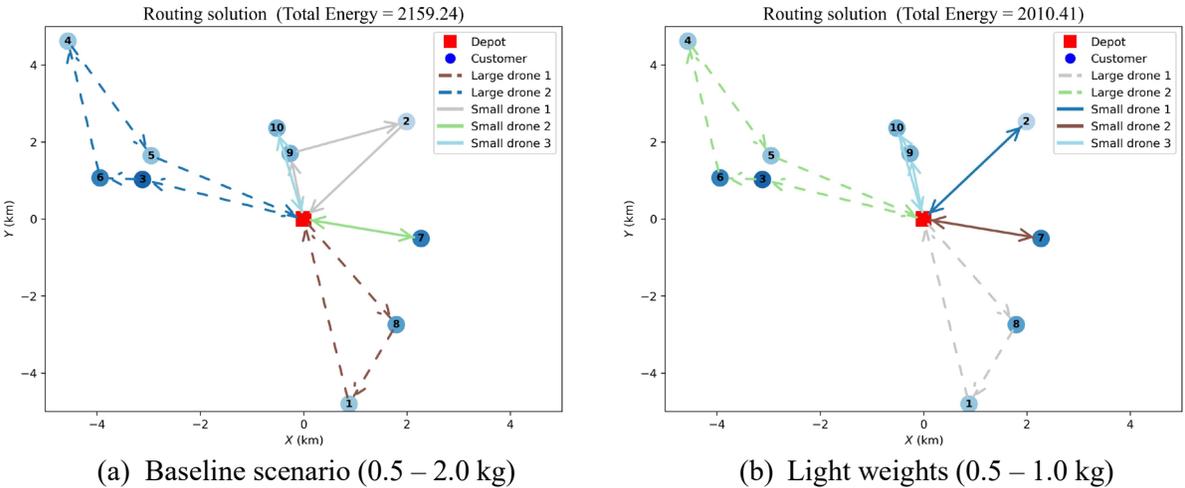

(a) Baseline scenario (0.5 – 2.0 kg)          (b) Light weights (0.5 – 1.0 kg)

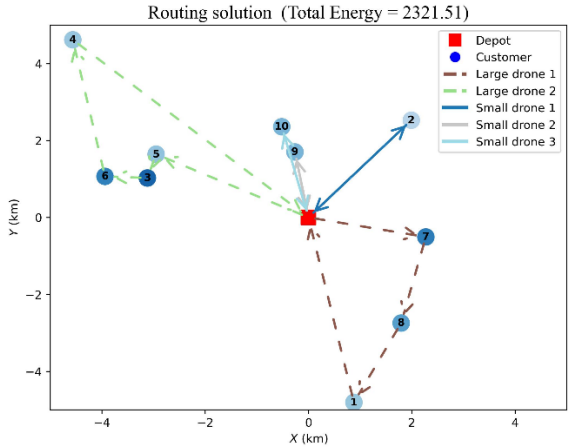
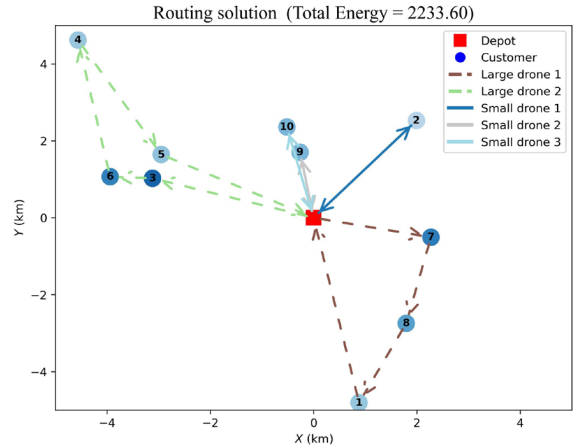

(c) Heavy weights (1.5 – 2.0 kg)   (d) Dispersed weights (0.25 – 3.0 kg)

**Figure 5 Sensitivity analysis of optimal solutions with respect to package weights.**

Fourth, we analyze the impact of the service area size by rescaling the customer locations to fit three scenarios: a small (5 km × 5 km), a large (15 km × 15 km), and a rectangular (5 km × 10 km) area. The results are shown in **Figure 6**.

The optimal routing strategies change significantly with the service area's dimensions. In the small 5 km × 5 km area (b), the roles of the drones shift: many small drones are now used for multi-stop tours (e.g., small drone 2 executes the tour 0 → 10 → 4 → 0). This demonstrates that for shorter overall distances, the lower fixed energy cost of small drones makes them more efficient. Conversely, in the large 15 km × 15 km area (c), the roles reverse. All small drones are restricted to single-customer deliveries due to their limited range, while large drones handle all multi-stop tours. This illustrates that as distance increase, the travel range and level-flight efficiency of large drones become critical, making them the better option for longer tours.

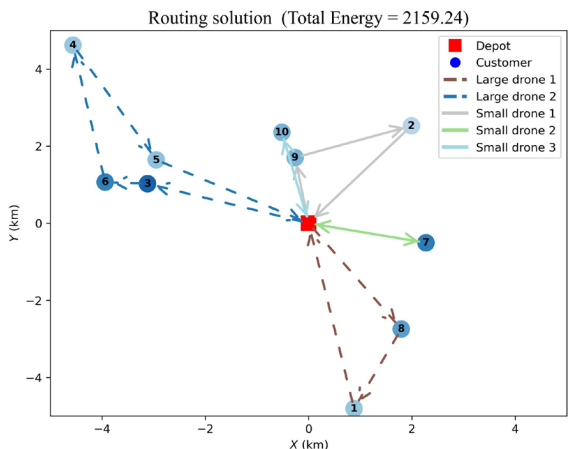
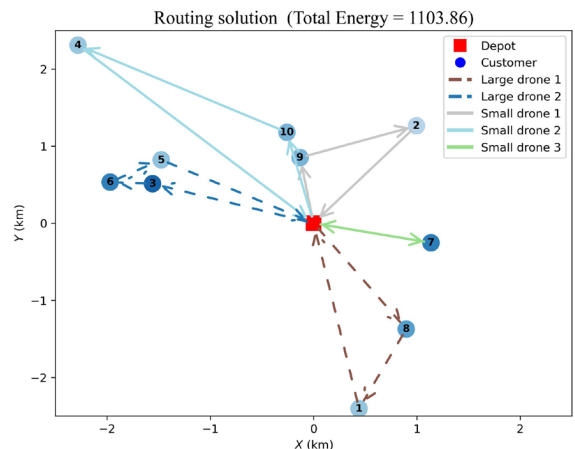

(a) Baseline scenario (10 km × 10 km)   (b) Small area (5 km × 5 km)

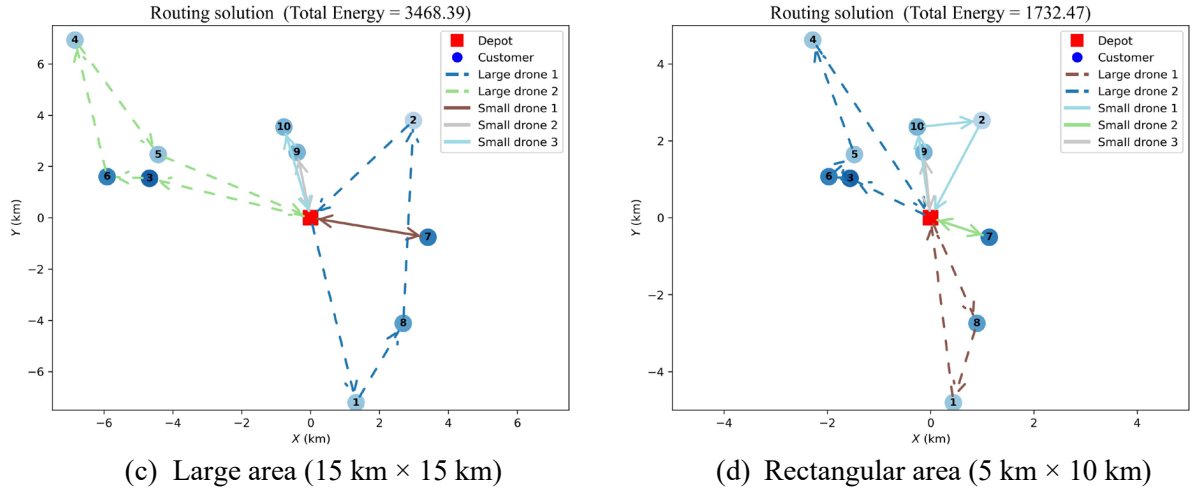

(c)  Large area (15 km × 15 km)    (d)  Rectangular area (5 km × 10 km)

**Figure 6 Sensitivity analysis of optimal solutions with respect to service area size.**

Finally, we test the scalability of the model by iteratively adding a new, randomly generated customer and re-solving the problem. The relationship between the number of customers, the objective value (total energy), and the required solution time is shown in **Table 4**.

**Table 4 Results and performance for different numbers of customers.**

| # Customers | 10 | 11 | 12 | 13 | 14 | 15 | 16 | 17 | 18 |
|---|---|---|---|---|---|---|---|---|---|
| Best MIP sol | 2159.24 | 2206.04 | 2534.44 | 2733.73 | 2749.91 | 2933.05 | 3111.72 | 3202.29 | 3284.64 |
| Time (s) | 3 | 10 | 25 | 102 | 394 | 243 | 709 | 1788 | 5000 |
| Rel. gap | 0 | 0 | 0 | 0 | 0 | 0 | 0 | 0 | 0.038 |

**Note:** The maximum solving time is capped at 5000 seconds. The first row is the number of customers. The second row is the best-found MIP solution. The third row is the solution time. The last row is the relative gap calculated as |*BestFound-BestPossible*|/*BestPossible*, which is reported by GAMS.

As expected, both the objective value and the solution time generally increase with the number of customers. The NP-hard nature of the problem is evident, as the solution time grows significantly, and an optimal solution for 18 customers could not be found within the 5,000-second time limit.

Interestingly, the solution time does not increase monotonically; for example, the time required to solve the 15-customer instance is less than that for the 14-customer instance. This can occur in branch-and-cut algorithms. The addition of a new customer can sometimes create a more constrained problem, where the tighter weight and energy limits prune the search tree more effectively, leading to a faster solution.

The model can find the optimal solution for problems with up to 17 customers in under 30 minutes. However, as the number of customers increases, the solution time increases significantly, highlighting the need for more efficient algorithms to solve large-scale instances. The primary objective of this paper is to understand *how* green routing strategies reduce energy consumption, not to develop a highly efficient solution algorithm. Therefore, the development of such an algorithm is a promising direction for future research.

**4.2 Solomon instances**

To ensure robust results, we also conduct numerical experiments using the Solomon dataset, which is widely used in truck-drone routing research (e.g., *Kang and Lee, 2021*; *Lu et al., 2025*). The nodes in the

1    Solomon data have three cases of distributions, named clustered (C), randomly generated (R), and semi-
2    clustered (RC). We select the first dataset from each distribution and rescale the node locations to fit a 5
3    km × 5 km area. For the payload weights, we rescale the original demand values into a range of 0.5 – 2.0
4    kg to maintain consistency with Section 4.1. From each 100-customer dataset, we create 10 subsets with
5    $n = 10$ customers each. All other parameters remain consistent with those listed in Table 3.
6
7    We present a comparison of the total energy consumption between drone delivery using minimum distance
8    routing and minimum energy routing. The results are summarized in Table 5. Across the 30 instances tested,
9    20 demonstrate energy savings when applying the green drone routing strategy. For the instances where
10   energy consumption is reduced, the average saving is 2.17%. Notably, in instance C101-1, the energy
11   saving reaches 5.97% compared to the minimum distance routing approach. These experiments further
12   demonstrate the potential of applying green drone routing strategies to reduce GHG emissions.
13
14   **Table 5 Computational results for Solomon dataset.**

| Problem | Total energy for minimum distance routing | Total energy for minimum energy routing (green drone routing) | Energy saving |
|---|---|---|---|
| C101-1 | 1345.27 | 1264.96 | 5.97% |
| C101-2 | 1245.29 | 1245.29 | 0 |
| C101-3 | 1033.26 | 1033.26 | 0 |
| C101-4 | 1202.67 | 1148.24 | 4.53% |
| C101-5 | 1104.82 | 1092.59 | 1.11% |
| C101-6 | 1345.85 | 1338.53 | 0.54% |
| C101-7 | 1126.46 | 1126.46 | 0 |
| C101-8 | 1127.42 | 1127.42 | 0 |
| C101-9 | 1178.78 | 1178.78 | 0 |
| C101-10 | 1185.53 | 1181.71 | 0.32% |
| R101-1 | 1594.99 | 1593.52 | 0.09% |
| R101-2 | 1510.85 | 1491.17 | 1.30% |
| R101-3 | 1423.21 | 1396.49 | 1.88% |
| R101-4 | 1425.25 | 1425.25 | 0 |
| R101-5 | 1428.19 | 1417.17 | 0.77% |
| R101-6 | 1281.45 | 1259.70 | 1.70% |
| R101-7 | 1466.37 | 1466.37 | 0 |
| R101-8 | 1047.25 | 1047.25 | 0 |
| R101-9 | 1314.30 | 1264.83 | 3.76% |
| R101-10 | 909.34 | 871.11 | 4.20% |
| RC101-1 | 1351.99 | 1313.13 | 2.87% |
| RC101-2 | 1417.12 | 1402.63 | 1.02% |
| RC101-3 | 1364.43 | 1297.48 | 4.91% |
| RC101-4 | 1301.09 | 1301.09 | 0 |
| RC101-5 | 1256.87 | 1241.48 | 1.22% |
| RC101-6 | 1651.85 | 1651.31 | 0.03% |
| RC101-7 | 1491.17 | 1429.38 | 4.14% |
| RC101-8 | 1555.72 | 1513.19 | 2.73% |
| RC101-9 | 1340.83 | 1340.83 | 0 |
| RC101-10 | 1170.72 | 1166.56 | 0.36% |

15
16
17

# 5. Conclusions

This paper presents a novel green drone routing problem (G-DRP) for a multi-visit, heterogeneous drone fleet with the objective of minimizing total energy consumption. Our analysis demonstrates that green drone routing strategies are fundamentally different from traditional distance-based vehicle routing. Specifically, we illustrate that the significant impact of payload on a drone's energy consumption leads to several counter-intuitive strategies: a longer route may be more energy-efficient if it allows for the early delivery of heavy packages; separate, single-customer tours can consume less energy than one multi-stop tour; and a heterogeneous fleet that matches drone size to the specific delivery task outperforms a homogeneous fleet. These findings underscore that for drone delivery, routing decisions must be driven by a holistic view of distance, payload, and drone type, not just travel distance alone. The numerical study confirms these insights; across the 30 instances tested, 20 demonstrate energy savings when applying the green drone routing strategy. For the instances where energy consumption is reduced, the average saving is 2.17% with a maximum saving of 5.97%, demonstrating significant potential for reducing the environmental footprint of drone delivery in last-mile logistics.

For future studies, several potential directions can be explored.

First, a reliable green drone routing solution requires an accurate energy consumption model. As discussed in **Appendix A,** to the best of our knowledge, there appears to be a lack of studies on examining drone energy consumption during the takeoff and landing phases. Future field studies are needed to provide a comprehensive evaluation of energy use across all four phases of flight.

Second, the computational time of our proposed green drone routing model increases significantly with the number of customers. Future work could develop more scalable solution methods, including both approximation algorithms (such as heuristics and metaheuristics) and more advanced exact methods (such as Benders decomposition and column generation).

Third, our model focuses on drone-only delivery. In practice, drones often operate in conjunction with other delivery modes, such as trucks or vans. Extending the model to consider a combined truck-drone delivery system under an energy minimization framework could offer new insights into strategic trade-offs between different transportation modes. Such integration could help identify when and how to deploy drones most effectively to reduce total energy consumption and environmental impacts.

Fourth, we consider only the scenario in which the delivery requirements for all customers are known in advance. In practice, however, last-mile delivery by drones begins from a depot, and the arrival times of packages at the depot may be uncertain. An interesting extension would be to consider continually arriving delivery requests, where a key decision is whether a drone should wait at the depot until enough packages accumulate to justify dispatch, or depart immediately with a partial load. Under this setting, the objective function could also incorporate minimizing customer waiting times in addition to energy consumption minimization.

Finally, the current model assumes a deterministic environment, where all parameters, such as weather conditions, package weights, and battery performance, are known with certainty. However, real-world operations are inherently uncertain. Incorporating these uncertainties into the model would improve its robustness and practical applicability. This could be achieved through a robust optimization or stochastic programming formulation.

## Acknowledgements

The authors Z. Li and Q. Guo were partially supported by National Natural Science Foundation (Grant No. 2400153).

## Authors' contributions

Ziyue Li: Conceptualization, Methodology, Data Curation, Formal Analysis, Writing, Original Draft;
Qianwen Guo: Supervision, Funding Acquisition, Project Administration, Writing, Review & Editing.
Paul Schonfeld: Writing, Review & Editing.

## Funding

The authors Z. Li and Q. Guo were partially supported by National Natural Science Foundation (Grant No. 2400153).

## Data availability

All data utilized in this research is public information.

## Competing interests

No potential conflict of interest was reported by the author(s).

## Appendix A. Drone energy consumption models.

A systematic review of drone energy consumption models was provided by Zhang et al. (*2021*). This subsection will focus on two aspects: the different phases of energy consumption for a drone flight and the energy consumption models widely used in the literature.

A drone's energy consumption should be computed differently from that of ground vehicles, because a drone flight consists of four distinct phases: takeoff, level flight, hovering, and landing (*Kirschstein, 2020*), as illustrated in **Figure A1**. A complete energy model for these phases must account for the power required to overcome air drag, provide lift, facilitate climbing, and supply internal electronics (*Langelaan et al., 2017*).

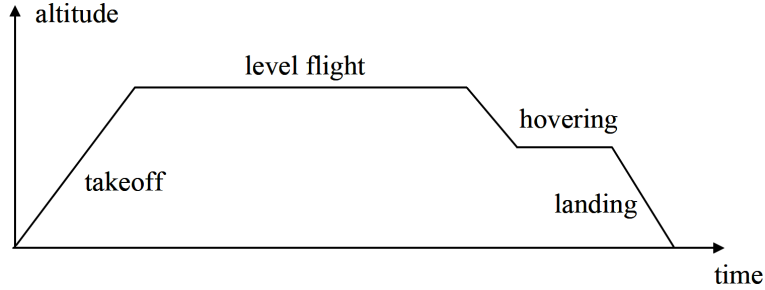

**Figure A1 Drone delivery flight pattern illustration. The figure is adapted from Kirschstein (*2020*).**

Figliozzi (*2017*) proposed an energy consumption model specifically for the steady level flight phase. The model provides an approximation of the required power based on the drone's total weight, speed, and lift-to-drag ratio. The power is formulated as:

$$P_1 = \frac{(m_t + m_b + m_l)gs}{r(s)},$$ (A1)

where $m_t, m_b, m_l$ are drone mass without battery and load, drone battery mass, and drone load mass, respectively. $g$ is gravitational acceleration. $r(s)$ is lift-to-drag ratio, which is a function of speed $s$.

The energy required to travel a given distance is the power multiplied by the travel time, and adjusted for the power transfer efficiency from the battery to the propellers (energy loss):

$$\frac{P_1 t}{\eta_p} = \frac{(m_t + m_b + m_l)gst}{r(s)\eta_p} = \frac{(m_t + m_b + m_l)gd}{r(s)\eta_p},$$ (A2)

where $d$ is the travel distance $= st$, and $\eta_p$ is the power transfer efficiency. While the travel speed $s$ drops out from the numerator, the lift-to-drag ratio $r(s)$ in the denominator is still dependent on speed. Therefore, to minimize energy consumption, a drone should fly at the speed that maximizes its lift-to-drag ratio. This optimal ratio is defined as $r^*$. Different aircraft types have different optimal lift-to-drag ratios. For instance, airplanes have $r^*$ values in the range of 10-20, and helicopters have values between 3.5-5.0 (*Leishman, 2006*). In contrast, the values for multirotor drones can be as low as 1 (*Gong et al., 2023*). Table 5 in Zhang et al. (*2021*) provides a summary of different models used to estimate drone energy consumption during steady level flight.

For the hovering phase, Dorling et al. (*2016*) provided a physics-based model for power consumption. The required power is:

$$P_2 = (m_t + m_b + m_l)^{\frac{3}{2}}\sqrt{\frac{g^3}{2\rho\zeta h}},$$ (A3)

where $\rho$ is the fluid density of air, $\zeta$ is the area of spinning blade disc, and $h$ is the number of rotors. Dorling et al. (*2016*) also proposed a linear approximation:

$$P_3 = \alpha(m_t + m_b + m_l) + \beta, \tag{A4}$$

where $\alpha$ and $\beta$ are two constant numbers obtained by a linear approximation. A similar linear model was also developed by Jeong et al. (*2019*).

Another similar model for hover power was provided by Liu et al. (*2017*):

$$P_4 = c_p[(m_t + m_b + m_l)g]^{\frac{3}{2}}, \tag{A5}$$

where $c_p$ is a parameter that depends on the details of the drone and the environmental parameters.

Unfortunately, we did not identify specific energy consumption models for the takeoff and landing phases. We therefore assume that the energy consumed during these phases can be approximated by a linear model, similar to Eq. (A4) used for hovering.[4]

Many studies (e.g., *Bruni et al., 2023*; *Cheng et al., 2020*; *Jiang et al., 2025*; *Masmoudi et al., 2022*; *Xia et al., 2023*) simplify their energy calculations by applying a hover power formula, such as those in Eqs. (A3) - (A5), to the entire flight. Cheng et al. (*2020*) explained that this approach was applied due to a lack of available field tests of small drones making multiple deliveries at the time.

Therefore, we adopt a linear model for our study. This approach is not only justified by existing studies, but it also allows energy consumption computation for all four flight phases to be consolidated into a single, tractable form.

**Appendix B. G-DRP reformulation.**

The objective function (5) is non-linear because it contains the product of a binary variable ($x_{ijkt}$) and a continuous variable ($w_{ikt}$). To formulate this into a MILP, we introduce a new continuous variable $z_{ijkt}$ to replace this product, i.e., $z_{ijkt} = x_{ijkt}w_{ikt}$. The objective function can now be written in linear form as:

$$\min \quad E = \sum_{i=0}^{n}\sum_{j=0}^{n}\sum_{k=1}^{K}\sum_{t=1}^{h_k}(e_{lk} + e_{fk}d_{ij})z_{ijkt}. \tag{B1}$$

This substitution requires four additional constraints to express the relationship $z_{ijkt} = x_{ijkt}w_{ikt}$:

$$z_{ijkt} \le W_k x_{ijkt}, \qquad \forall i,j \in N \cup \{0\}, k \in \mathcal{K}, t \in H_k, \tag{B2}$$

$$z_{ijkt} \ge m_{0k} x_{ijkt}, \qquad \forall i,j \in N \cup \{0\}, k \in \mathcal{K}, t \in H_k, \tag{B3}$$

$$z_{ijkt} \le w_{ikt} - m_{0k}(1 - x_{ijkt}), \qquad \forall i,j \in N \cup \{0\}, k \in \mathcal{K}, t \in H_k, \tag{B4}$$

$$z_{ijkt} \ge w_{ikt} - W_k(1 - x_{ijkt}), \qquad \forall i,j \in N \cup \{0\}, k \in \mathcal{K}, t \in H_k. \tag{B5}$$

These constraints work as follows. If $x_{ijkt} = 0$, constraints (B2) and (B3) force $z_{ijkt} = 0$. Constraints (B4) and (B5) are also satisfied in this case because the drone's total weight ($w_{ikt}$) will always be within its operational bounds, i.e., $m_{0k} \le w_{ikt} \le W_k$. If $x_{ijkt} = 1$, constraints (B4) and (B5) enforce $z_{ijkt} = w_{ikt}$. Constraints (B2) and (B3) are also naturally satisfied in this case.

Similarly, the non-linear energy capacity constraint (15) is reformulated by substituting the product of variables with $z_{ijkt}$. This results in the following linear constraint:

---

[4] This assumption is reasonable because the energy required for takeoff is dominated by climbing against gravity, which can be approximated by the change in gravitational potential energy and is therefore proportional to the mass.

$$\sum_{i=0}^{n}\sum_{j=0}^{n}\left(e_{lk} + e_{fk}d_{ij}\right)z_{ijkt} \leq E_k, \qquad \forall k \in \mathcal{K}, t \in H_k.$$ (B6)

The non-linear weight update rule (12) can be replaced with a standard "big-M" linearization. This requires the following two linear constraints:

$$w_{ikt} - w_{jkt} - m_j \leq \left(W_k - m_{0k} - m_j\right)\left(1 - x_{ijkt}\right), \qquad \forall i \in N \cup \{0\}, j \in N, \forall k \in \mathcal{K}, t \in H_k,$$ (B7)

$$w_{ikt} - w_{jkt} - m_j \geq \left(m_{0k} - W_k - m_j\right)\left(1 - x_{ijkt}\right), \qquad \forall i \in N \cup \{0\}, j \in N, \forall k \in \mathcal{K}, t \in H_k.$$ (B8)

These constraints ensure the correct weight update for two possible cases. If the drone travels from $i$ to $j$ ($x_{ijkt} = 1$), the right-hand side of both inequalities becomes zero, which forces $w_{jkt} = w_{ikt} - m_j$. If no travel occurs ($x_{ijkt} = 0$), the constraints are redundant, as the difference in weights between any two points on a tour will not exceed the drone's maximum possible weight change, i.e., $|w_{ikt} - w_{jkt}| \leq |W_k - m_{0k}|$.

Finally, the subtour elimination constraint (13) contains a product of decision variables. To linearize this, we introduce a new variable $s_{ijkt} = x_{ijkt} \sum_{i'=0}^{n} \sum_{j'=1}^{n} x_{i'j'kt}$ to represent this product. The original constraint is then replaced by the following set of linear constraints:

$$u_{ikt} - u_{jkt} + s_{ijkt} \leq \sum_{i'=0}^{n} \sum_{j'=1}^{n} x_{i'j'kt} - 1, \qquad \forall i, j \in N, \forall k \in \mathcal{K}, t \in H_k,$$ (B9)

$$s_{ijkt} \leq n x_{ijkt}, \qquad \forall i, j \in N, \forall k \in \mathcal{K}, t \in H_k,$$ (B10)

$$s_{ijkt} \leq \sum_{i'=0}^{n} \sum_{j'=1}^{n} x_{i'j'kt}, \qquad \forall i, j \in N, \forall k \in \mathcal{K}, t \in H_k,$$ (B11)

$$s_{ijkt} \geq \sum_{i'=0}^{n} \sum_{j'=1}^{n} x_{i'j'kt} - n\left(1 - x_{ijkt}\right), \qquad \forall i, j \in N, \forall k \in \mathcal{K}, t \in H_k,$$ (B12)

$$s_{ijkt} \geq 0, \qquad \forall i, j \in N, \forall k \in \mathcal{K}, t \in H_k.$$ (B13)

Constraints (B9) – (B13) enforce the desired relationship. If $x_{ijkt} = 0$, constraints (B10) and (B13) force $s_{ijkt} = 0$. If $x_{ijkt} = 1$, constraints (B11) and (B12) force $s_{ijkt}$ to equal the total number of customers served on that tour (i.e., $s_{ijkt} = \sum_{i'=0}^{n} \sum_{j'=1}^{n} x_{i'j'kt}$). This completes the reformulation of the G-DRP into a MILP.

A key challenge in this formulation is the presence of symmetrical solutions. For any given drone type $k$, if $\tilde{h}_k$ drones are dispatched out of the $h_k$ available, there are $\binom{h_k}{\tilde{h}_k}$ ways to choose which specific drones are used. Furthermore, there are $\tilde{h}_k!$ ways to assign the same set of routes to these dispatched drones. This creates a large number of equivalent solutions, which can significantly slow down the solution process.

To address this, the following valid inequalities, adapted from Adulyasak et al. (*2014*), can be added to the model to break the symmetry. First, we can force the drone tours for each type to be used in a sequential order. The following constraint allows tour $t + 1$ of a given type to be dispatched only if tour $t$ is also dispatched:

$$\sum_{j=1}^{n} x_{0jkt} \geq \sum_{j=1}^{n} x_{0,j,k,t+1}, \qquad \forall k \in \mathcal{K}, t \in H_k \setminus \{h_k\}.$$ (B14)

Second, we can apply lexicographical ordering constraints to ensure that for any two tours of the same drone type, the tour with the smaller index ($t$) serves the "smaller" set of customers. We introduce a binary variable $y_{ikt}$ that is 1 if customer $i$ is served by drone $k$ on tour $t$. This is linked to the original variables, and the ordering is enforced as follows:

$$\sum_{i=1}^{n} 2^{n-i} y_{ikt} \geq \sum_{i=1}^{n} 2^{n-i} y_{i,k,t+1}, \qquad \forall k \in \mathcal{K}, t \in H_k \backslash \{h_k\}, \tag{B15}$$

$$y_{ikt} = \sum_{j=0}^{n} x_{ijkt}, \qquad \forall i \in N, k \in \mathcal{K}, t \in H_k. \tag{B16}$$

## Appendix C. Extensions.

### C.1 No-fly zones

Many countries have strict regulations on drone use due to safety and privacy concerns, often designating certain areas as no-fly zones (*Kang & Lee, 2021*). In the United States, the Federal Aviation Administration (FAA) defines three main types of no-fly zones: restricted airspace, where all drone flights are prohibited; local restrictions, imposed by state, local, territorial, or tribal government agencies on takeoffs and landings; and temporary flight restrictions, which prohibit drone operations in specific areas for limited periods, such as during major sporting events or presidential movements. These restrictions vary in size, altitude, date/time, and types of operations (*Federal Aviation Administration, 2025*). Similar regulations exist in other countries, including the United Kingdom and France (*Mandourah & Hochmair, 2024*).

Incorporating no-fly zones into the G-DRP model presents two cases. First, if a customer is located inside a no-fly zone, the delivery problem for that customer becomes infeasible. Second, if a customer is outside the zone but the most flight path intersects it, one can adjust the distance matrix, $d_{ij}$, to reflect the increased travel distance required to fly around the restricted area.

### C.2 Volume limit for payload

In addition to weight, drones are also constrained by the physical volume of their payload compartment. We can model this space limit $V_k$ for each drone type. Assuming each package for customer $i$ has a volume of $v_i$, the following constraint can be added to the G-DRP model:

$$\sum_{i=1}^{n} \sum_{j=0}^{n} v_i x_{ijkt} \leq V_k, \qquad \forall k \in \mathcal{K}, t \in H_k. \tag{C1}$$

This constraint ensures that the total volume of all packages assigned to any single drone tour does not exceed that drone's space limit.[5]

### C.3 Weather conditions

Weather conditions, particularly wind, can significantly affect a drone's energy consumption and flight speed. There are two primary ways to incorporate these effects into the model. First, for general weather conditions (e.g., high temperature, rain), the drone's optimal lift-to-drag ratio ($v^*$) can be recalculated. This would in turn adjust the energy consumption coefficients ($e_{lk}$ and $e_{fk}$) for each drone type. Second, if specific wind speed and direction are known, the energy parameters can vary for different flight legs. For each flight leg from node $i$ to $j$, unique energy coefficients, $e_{lk}^{ij}$ and $e_{fk}^{ij}$, that account for the effects of headwinds, tailwinds, or crosswinds, can be computed. These specific values would then replace the global $e_{lk}$ and $e_{fk}$ parameters in the optimization model to provide a more accurate energy calculation.

---

[5] This is a simplification. In reality, the varied shapes and dimensions of packages may prevent the full volume of the payload compartment from being fully utilized. A more detailed model would require a container loading formulation (*Chen et al., 1995*), which is beyond the scope of this paper.

## C.4 Battery replacement

When accounting for the possibility of replacing batteries at the depot, the model can be reinterpreted. In this case, a single drone could perform multiple tours with freshly swapped batteries, the parameter $h_k$ would no longer represent the number of available drones of type $k$, but rather the total number of available tours (or equivalently, battery sets) for that drone type.

## C.5 Delivery with time window

To incorporate delivery time windows, we introduce a new continuous variable, $r_{ikt}$, which represents the service start time at customer $i$ by drone $k$ on tour $t$. Each customer $i$ has a required time window $[a_i, b_i]$. Let $\tau_{ijk}$ be the travel time from node $i$ to $j$ for drone $k$, and $\tau_0$ be the service time required at each customer location.

With the introduction of the service time variable, the MTZ subtour elimination constraint (13) is no longer needed and can be replaced by the following constraints:

$$x_{ijkt}(r_{ikt} + \tau_0 + \tau_{ijk} - r_{jkt}) \leq 0, \qquad \forall i,j \in N, k \in \mathcal{K}, t \in H_k, \tag{C2}$$
$$a_i \leq r_{ikt} \leq b_i, \qquad \forall i \in N, k \in \mathcal{K}, t \in H_k. \tag{C3}$$

Constraint (C2) establishes the sequence of deliveries. If a drone travels from $i$ to $j$, the service at $j$ must begin after the service at $i$ is complete and the drone has traveled to $j$. Constraint (C3) ensures that the service time at each customer falls within that customer's specified time window. Together, these time-based constraints inherently prevent subtours. If a particular drone tour does not serve customer $i$, the value of $r_{ikt}$ becomes irrelevant to the solution (*Kallehauge et al., 2005*).

## Appendix D. Details of customer instances generated in Section 4.1.

*Customer location*

| Customer | 1 | 2 | 3 | 4 | 5 | 6 |
|----------|---|---|---|---|---|---|
| Location | (0.88, -4.81) | (1.99, 2.53) | (-3.12, 1.02) | (-4.56, 4.62) | (-2.95, 1.64) | (-3.94, 1.07) |
| Customer | 7 | 8 | 9 | 10 | 11 | 12 |
| Location | (2.27, -0.51) | (1.79, -2.75) | (-0.26, 1.70) | (-0.52, 2.36) | (-2.14, 0.05) | (2.40, 3.49) |
| Customer | 13 | 14 | 15 | 16 | 17 | 18 |
| Location | (-2.61, -2.06) | (-0.62, 1.77) | (3.84, -0.79) | (-2.11, 1.82) | (2.85, -2.79) | (2.59, 0.49) |

*Goods weight*

| Customer | 1 | 2 | 3 | 4 | 5 | 6 | 7 | 8 | 9 |
|----------|---|---|---|---|---|---|---|---|---|
| Weight | 0.89 | 0.64 | 1.94 | 0.88 | 0.92 | 1.65 | 1.70 | 1.32 | 1.07 |
| Customer | 10 | 11 | 12 | 13 | 14 | 15 | 16 | 17 | 18 |
| Weight | 1.07 | 1.12 | 1.67 | 1.91 | 0.66 | 1.91 | 1.70 | 1.00 | 0.97 |